# One-Dimensional Metallic Polymeric Nitrogen

Kewei Ding,*,a,b Junyi Miao,[c] Ying Liu,[a,d] Anxin Yu,[a] Cheng Lu,*,c Wenrui Zhang,[c] Yanchun Li,*,e Haipeng Su,[a] Zhongxue Ge,[a] and Xianlong Wang,*,f

[a]Xi'an Modern Chemistry Research Institute, Xi'an 710065, China. E-mail: dkw204@163.cn

[b]State Key Laboratory of Fluorine & Nitrogen Chemicals, Xi'an 710065, China

[c]School of Mathematics and Physics, China University of Geosciences (Wuhan), Wuhan 430074, China. E-mail: lucheng@calypso.cn

[d]Quantum Science Center of Guangdong-Hong Kong-Macao Greater Bay Area (Guangdong), Shenzhen 518045, China.

[e]Institute of High Energy Physics, Chinese Academy of Sciences, Beijing 100049, China. E-mail: liyc@ihep.ac.cn

[f] Key Laboratory of Materials Physics, Institute of Solid State Physics, HFIPS, Chinese Academy of Sciences, Hefei 230031, China. E-mail: xlwang@theory.issp.ac.cn

**Abstract**

**The pressure-induced metallic states of light elements attract significant attention, because of potential applications as high-temperature superconductor and high-energy-density material, especially for hydrogen and nitrogen[1-10]. Several semiconducting polymeric nitrogen phases with three- or two-dimensional sp³-bonded networks were synthesized[6-10], but its metallic form remains unobserved.**

**Here, we report the synthesis of a metallic polymeric nitrogen with one-dimensional feature (1D-PN) at 130–140 GPa and >3000 K. Synchrotron XRD and Raman spectroscopy, supported by DFT calculations, reveal that it adopts an infinite arm-chair like chain with $sp^2$-hybridized π-bonds. Simulations predict a superconducting transition at 21.19 K under 113 GPa, higher than that reported in high-pressure experiments for non-metallic elements. At ambient pressure, this phase acquiring an energy density of as high as 8.78 kJ/g is not only kinetically stable but also thermodynamically more stable than cubic gauche nitrogen. This multifunctional property profile positions 1D-PN as a disruptive candidate for both electronic and energetic applications.**

Nitrogen ($N_2$) and hydrogen ($H_2$) are two of the most extensively studied homonuclear diatomic molecules under high-pressure, owing to fundamental importance and intriguing characteristics[1-16]. The most important pursues in this field are the realization of atomic polymerized phase and metallization, since the atomic polymerized nitrogen[6-10] and hydrogen[1-4] are believed to be the revolutionary high-energy-density materials. Furthermore, metallic properties and small atomic mass can lead to a high superconducting transition temperature, e.g., metallic hydrogen is believed to exhibit room-temperature superconductivity[17]. However, due to limitations in ultra-high-pressure generation and in situ measurement techniques, more evidence is needed to confirm the polymerization and metallization of hydrogen, which needs a pressure higher than 500 GPa[1-4,18]. Polymeric nitrogen had been realized under relatively lower pressures of >110 GPa[6-10,19,20], while the metallic nitrogen with atomic polymerization has not been observed, similar to the case for other homonuclear diatomic molecules.

Different from hydrogen with only 1s electron, nitrogen's electronic configuration ($2s^22p^3$) enables the formation of diverse allotropes at high-pressure resulting in a complex high-pressure and high-temperature phase diagram[21], which provides a good reference for the pressure-induced structural evolution of other homonuclear diatomic molecules. Since 1980s, extensive theoretical reports have proposed numerous polymeric nitrogen structures[22-28]. However, only a few of these predicted phases have been experimentally synthesized[6-10,19,20]. The cubic gauche nitrogen (cg-N, the ground sate from 64 GPa to 190 GPa) with three-dimensional (3D) network of N-N single bond was first synthesized in 2004 at 110 GPa and 2000 K[6]. Following, three kinds of metastable polymeric nitrogen with two-dimensional (2D) structure were synthesized[7-10,20]. In 2014, a layered polymeric nitrogen (LP-N) composed of four atomic layers was experimentally obtained above 125 GPa and 2000 K coexisting with the cg-N and amorphous nitrogen[7]. In 2019, near 250 GPa, a hexagonal layered polymeric nitrogen (HLP-N) comprising four atomic layers was also identified[8]. Following that, a black phosphorus like nitrogen (BP-N) composed of two atomic layers was reported at 140 GPa and ~4000 K using Au as a laser absorber[9]. Meanwhile, the BP-N was synthesized along with cg-N at 146 GPa and 2200 K without additional thermal absorber[10,20].

To date, in all of synthesized polymeric nitrogen structures, the nitrogen atom has $sp^3$-type hybridization, which results in the semiconducting property[5,26]. For example, at 100 GPa, the cg-N with 3D-network, 2D LP-N composed of four atomic layers, and 2D BP-N composed of two atomic layers has a bandgap of 4.70 eV, 3.04 eV, and 1.87 eV (Supplementary Table 1), respectively[26]. We can find that the bandgap of synthesized polymeric nitrogen generally decreases as their dimensionality decreases. According to this scenario, a polymeric nitrogen structure with one-dimensional (1D) character is expected to exhibit a further reduced bandgap, potentially even demonstrating metallic behavior. This hypothesis is reasonable, since the $sp^2$-type hybridization can induce alternating single and double bonds along quasi-1D N-chain resulting in a π-conjugated electronic backbone. However, the polymeric nitrogen structure with 1D feature has not been synthesized yet.

Here, we report a one-dimensional metallic polymeric nitrogen phase (1D-PN)

with chain-like structure, synthesized in laser-heated diamond anvil cell (DAC) at pressures of 130–140 GPa and above 3000 K. The structure was identified through in situ synchrotron X-ray diffraction (XRD), Raman spectroscopy, and density functional theory (DFT) calculations. Simulations show that it has a superconducting transition temperature as high as 21.19 K at 113 GPa. The 1D-PN is kinetically stable at 0 GPa and can be potentially quenched to ambient pressure, where it has an energy density of 8.78 kJ/g and is thermodynamically more stable than cg-N. The realization of this metallic polymeric nitrogen phase at relatively moderate pressures not only provides a material that integrates high-energy-density with novel electronic functionality, but also establishes nitrogen as a versatile platform for studying metallization and potential superconductivity in light-element systems under high-pressure.

**Results and discussion**

High-pressure and high-temperature experiments were conducted using symmetric DACs equipped with a double-sided laser heating system. Polymeric nitrogen was synthesized by laser-heating high-purity $N_2$ at initial pressures of 130–140 GPa. For efficient heating, a bulk aluminum absorber (occupying ~1/3 of the chamber) served as the laser absorber alongside a NaCl thermal insulator. The sample was locally heated on multiple points with 60 W laser power, reaching temperatures above 3000 K. After quenching to room temperature, the pressure at the sample center relaxed to about 113 GPa. The resulting sample ($N_2$-Al-1) was then characterized by Raman spectroscopy and X-ray diffraction. As shown in Fig. 1a, the Raman spectrum revealed a new nitrogen allotrope characterized by four distinct peaks at 362.57, 454.50, 572.89, and 628.98 cm$^{-1}$, respectively. These spectral features are distinct from all known nitrogen phases, such as cg-N (single peak at 750-875 cm$^{-1}$) [6,19], LP-N (two colossal peaks at ~1000 and ~1300 cm$^{-1}$)[7], BP-N (two intense peaks at 860-1020 and 1200-1310 cm$^{-1}$)[9,10,20], HLP-N (twelve distinct vibrational modes at frequencies between 300 and 1300 cm$^{-1}$)[8], etc. The representative comprehensive XRD patterns are shown in Fig. 1b. As the laser-absorber Al was sandwiched between thermal insulator NaCl, diffraction peaks from both Al and NaCl were detected, along with those of the new phase formed after the laser-heating-induced reaction. Comparative experiments were also performed by either replacing or entirely removing the absorber material, where a higher laser power (~80 W) was used to compensate for the lack of absorption. In all cases, the resulting Raman spectra remained highly consistent (Supplementary Fig.1), confirming the product as a novel pure polymeric nitrogen phase. Detailed experimental information is summarized in Supplementary Table 2 for reference.

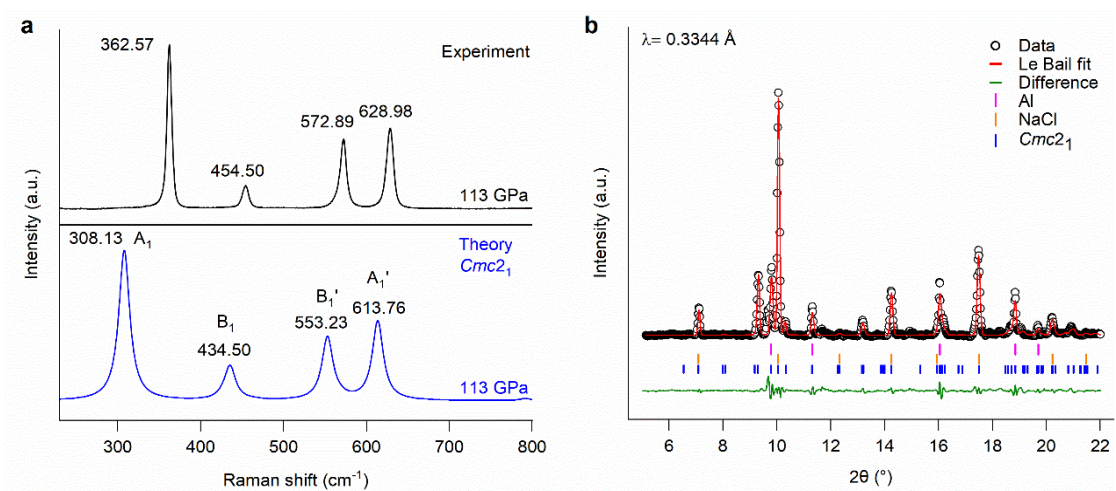

Fig. 1. Raman spectra and XRD patterns of the 1D-PN. **a**, Comparison of the experimental Raman spectrum (black) of laser-heated sample ($N_2$-Al-1) synthesized at 130 GPa and calculated Raman spectrum (blue) of 1D-PN with $Cmc2_1$ symmetry at 113 GPa. **b**, Le Bail refinement pattern of laser-heated sample ($N_2$-Al-1) measured at 113 GPa. The vertical ticks correspond to the Bragg peaks of Al (pink), NaCl (orange) and 1D-PN with $Cmc2_1$ symmetry at 113 GPa (blue).

To elucidate the crystal structure of the newly discovered polymeric nitrogen, a systematic structure searches were performed using the CALYPSO method[29-31]. The searched results confirmed the existence of multiple allotropes of nitrogen with distinct symmetries. For example, the molecular configurations ($Pa\bar{3}$ α phase, $P4_2/mnm$ γ phase, and $R\bar{3}c$ ε phase), and the polymeric geometries ($I2_13$ cg phase, $Cmce$ BP phase, $Pba2$ LP phase, $C2/c$ λ phase, and $P4_12_12$ θ phase)[22-28]. To explore the relative stabilities, enthalpy difference curves referenced to the cg-N are calculated. The structure phase diagram of nitrogen under high-pressure is depicted in Fig. 2a. The structural phase transition sequence of nitrogen under high-pressure from 0 GPa to 200 GPa are α-N, θ-N and cg-N. At ambient pressure, the molecular α-N phase exhibits the lowest energy structure. The cg-N phase dominates energetically between 64–190 GPa. The LP phase shows distinct energy superiority above 190 GPa. These results are in good agreement with previous findings[25-28], validating the accuracy and reliability of the current searched methods. Remarkably, a new orthorhombic nitrogen structure with $Cmc2_1$ symmetry was discovered. This polymeric phase contains 16 atoms per unit cell and exhibits clear bonding-reorganization characteristics under high-pressure. It becomes metastable above 19 GPa, but below this pressure it is significantly more stable than the well-known cg-N phase. Interestingly, the simulated Raman spectra of the *Cmc*$2_1$ phase under 113 GPa displays a characteristic vibrational profile, with major Raman peaks at 308.13, 434.50, 553.23, and 613.76 cm$^{-1}$. As shown in Fig. 1a, these simulated peaks closely match the experimental Raman peaks both in position and relative intensity, supporting the structural assignment.

To further confirm that the synthesized polymeric nitrogen corresponds to the *Cmc*$2_1$ phase, we compared the experimental XRD pattern with the simulated diffraction profile of this phase. Nitrogen diffraction under ultrahigh pressure is intrinsically weak and highly sensitive to non-hydrostatic stress and preferred orientation, making it difficult to obtain high-quality diffraction patterns. Despite these challenges, a three-phase Le Bail refinement was successfully performed on the XRD pattern of the laser-heated sample $N_2$-Al-1 (including Al and NaCl as secondary phases). The refinement yielded the lattice parameters of the polymeric nitrogen phase as a = 3.7056 Å, b = 4.7366 Å, and c = 4.1677 Å. Several new diffraction peaks (e.g., at 2θ = 9.3°, 10.3°, and 13.2°), which are not attributable to Al or NaCl, align well with the simulated Cmc$2_1$ pattern in terms of position, systematic absences, and other key features. The agreement between the experimental XRD pattern and theoretical predictions, combined with the matching Raman data, supports the assignment of the synthesized high-pressure polymeric nitrogen to the *Cmc*$2_1$ crystal structure.

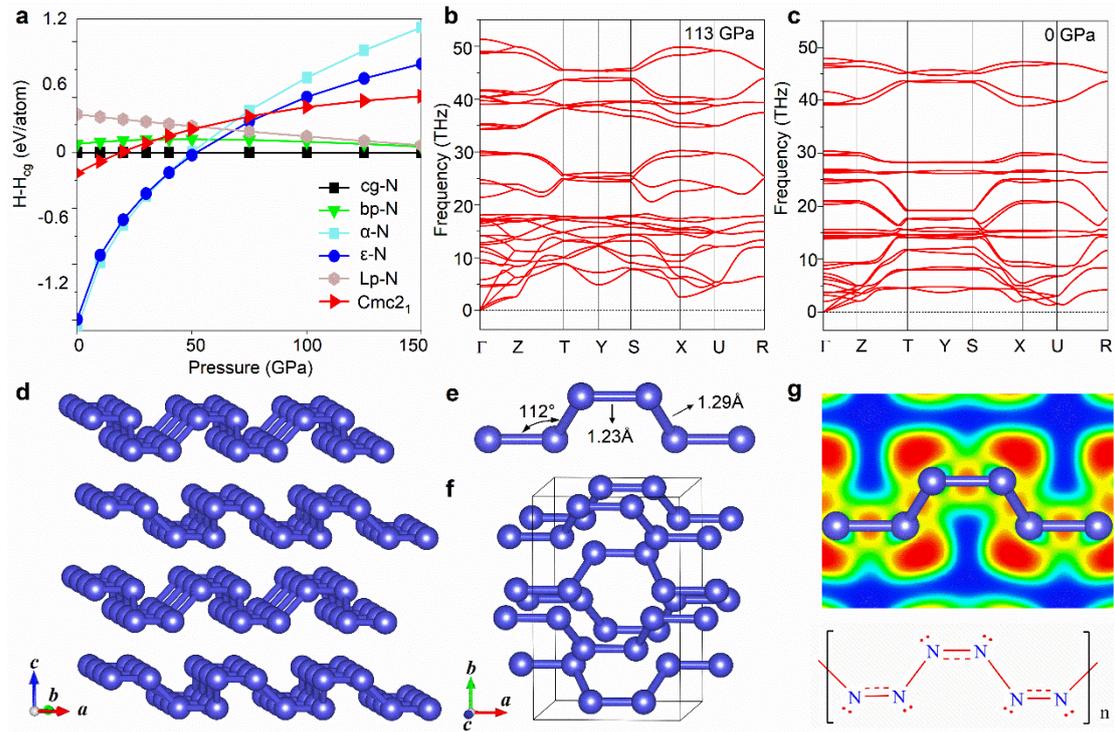

Fig. 2. **a**, Structural phase diagram of nitrogen at high-pressure (relative to the cg-N). **b**, The phonon dispersion curves of the $Cmc2_1$ phase at 113 GPa. **c**, The phonon dispersion curves of the $Cmc2_1$ phase at 0 GPa. **d**, The crystal structure of the $Cmc2_1$ phase at 113 GPa. **e**, Local chain geometry in the $Cmc2_1$ phase. **f**, View of the $Cmc2_1$ phase along (010). **g**, The ELF of the $Cmc2_1$ phase at 113 GPa. The Lewis structure of the nitrogen chain block is shown in the bottom.

To evaluate the kinetic stability of the orthorhombic $Cmc2_1$ phase, its phonon dispersion curves were calculated at both 113 GPa and 0 GPa, as shown in Fig. 2b and Fig. 2c. The absence of imaginary frequencies throughout the Brillouin zone indicates that the phase is dynamically stable across this entire pressure range. Furthermore, as summarized in Supplementary Table 3, the calculated independent elastic constants ($C_{ij}$) show that the $Cmc2_1$ phase is also mechanical stable at the pressure ranging from 0 GPa to 113 GPa.

The crystallographic and electronic structures of the *Cmc*2$_1$ phase of nitrogen were investigated, and corresponding results are shown in Fig. 2d-f. The crystal structure of *Cmc*2$_1$ phase under 113 GPa is presented in Fig. 2d, where a corrugated polymeric framework is built from saw-tooth N$_6$ units that fuse into infinite 1D chains running strictly along the crystallographic *a*-axis (red arrow). A fragment of a single chain of *Cmc*2$_1$ phase of nitrogen (1D-PN) is depicted in Fig. 2e. The two distinct N–N bond lengths are 1.23 Å and 1.29 Å, and the N–N–N angle is 112°, all within the typical range for N=N double bonds (1.20~1.35 Å). The packing of these chains is visualized in Fig. 2f (view along (010)), which reveals corrugated layers in the (010) plane. Nitrogen atoms occupy two distinct Wyckoff 8*b* sites: (0.17735, 0.80672, 0.85649) and (0.32027, 0.04111, 0.53299) (Supplementary Table 4). The electron localization function (ELF) and Lewis structures in Fig. 2g confirms the double-bond character, exhibiting a [-N=N-N=N-] bonding motif. The energy band and projected density of states (PDOS) of the *Cmc*2$_1$ phase under 113 GPa are presented in Supplementary Fig.2, which indicates that the metallic properties mainly originate from the p-orbital of nitrogen. The ELF and Lewis structures also verifies the presence of π-bonding connections between the adjacent nitrogen atoms, indicating the sp² hybridization in 1D-PN at 113 GPa.

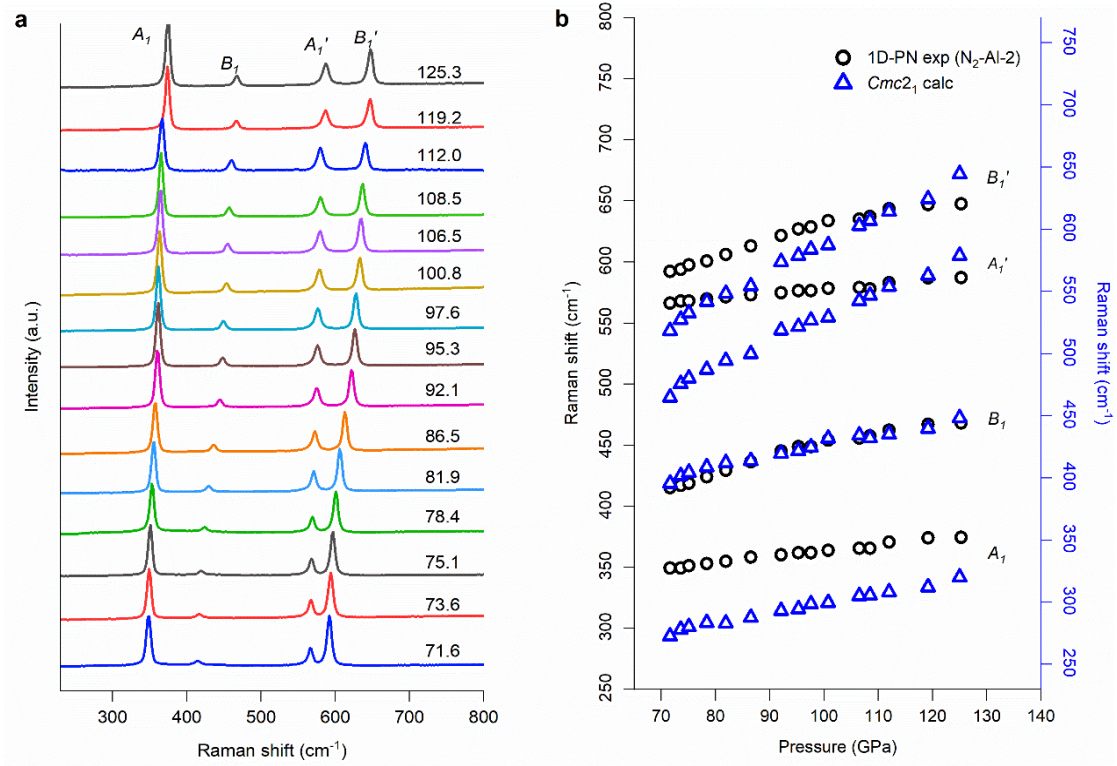

Fig.3. **a**, Raman spectra of the new polymeric nitrogen sample ($N_2$-Al-2) recorded during the decompression. **b**, Pressure dependence of the Raman peak positions. The experimental data are shown as black open circles (left axis), and the calculated values for the $Cmc2_1$ phase are plotted as blue triangles (right axis).

We synthesized an additional 1D-PN sample ($N_2$-Al-2) and investigated its stability during decompression. The Raman spectra recorded throughout this pressure release process are presented in Fig. 3a. For 1D-PN, the four calculated Raman peaks in the range of 300–650 cm$^{-1}$ are unambiguously assigned to the $A_1$, $B_1$, $A_1'$, and $B_1'$ phonon modes. As illustrated in Supplementary Fig. 3, their respective atomic vibration patterns are characterized as follows: the $A_1$ mode corresponds to torsional vibration about the chain axis, the $B_1$ mode to a wavelike buckling along the chains, the $A_1'$ mode to inter-chain sliding, and the $B_1'$ mode to out-of-plane wagging of the chains. As shown in Fig. 3a, upon pressure decrease, the Raman modes $A_1'$ and $B_1'$ gradually approach each other, while the intensity of the $A_1$ and $B_1$ modes slowly diminishes. The pressure could only be reduced to 71.6 GPa, below which sample escaped due to failure of a diamond anvil[32]. The pressure dependence of the Raman-mode frequencies during decompression is summarized in Fig. 3b, which compares the experimental data (open circles) with calculated results (blue triangles). The calculated frequencies of the $A_1$, $B_1$, and $B_1'$ modes are in broad agreement with the experimental values, while the pressure coefficient ($dv/dP$) of the $A_1'$ mode shows a discernible deviation. This discrepancy likely originates from the particular sensitivity of the inter-chain sliding vibration to anisotropic stress under non-hydrostatic conditions and possible inaccuracies in modeling weak inter-chain interactions[10].

The energy band structures of $Cmc2_1$ phase suggest that the 1D-PN is metallic and potential superconductor. The superconducting properties of 1D-PN under high-pressure are estimated using the McMillan equation derived from Bardeen–Cooper–Schrieffer (BCS) theory. The calculated superconducting transition temperature (Tc) of 1D-PN is 21.19 K under 113 GPa, making it the highest among experimentally realized superconductors of non-metallic elements under high-pressure (See Supplementary Table 5). Typically, for elemental sulfur, the maximum Tc value reaches 17 K at 157 GPa in the S-IV phase, and decreases to 15 K in the S-V phase above 155 GPa[33]. Elemental oxygen exhibits superconductivity with Tc=0.6 K in the ζ-phase at 115–120 GPa, which is classified as unconventional superconductivity[34]. As pressure increases from 0 GPa to 113 GPa, the Tc value increases monotonically, demonstrating a clear positive correlation between pressure and superconductivity in 1D-PN. Detailed analysis (see Supplementary Fig. 4) show that the softened phonon modes enhance the electron–phonon coupling (EPC) interactions and increase the Tc value to be 21.19 K under 113 GPa. The potential superconducting mechanism is mainly driven by strong coupling interactions induced by low-frequency soft phonons, which promotes Cooper pair formation and supports the increase of Tc values.

It is worth mentioning that 1D-PN exhibits excellent high-energy characteristics. The calculated results show that 1D-PN shows a remarkably high density of 2.70 g/cm³ at ambient pressure. Upon decomposition, it releases approximately 1.27 eV of chemical energy per atom, corresponding to an outstanding energy density of 8.78 kJ/g. The ultrahigh energy density and gas yield endow 1D-PN with obviously advantageous for explosive and propulsion applications. The predicted detonation velocity and specific impulse reach 15,721 m/s and 344.5 s, respectively, significantly higher than those of the current top-performing conventional energetic material of CL-20 (approximately 9,455 m/s and 251 s)[35]. Moreover, the metallic properties lead to the coupling of intense electromagnetic pulses during the detonation process, thereby giving rise to novel energy-release phenomena and achieving transformative explosive performance.

**Conclusions**

In summary, a novel one-dimensional polymeric nitrogen (1D-PN) was synthesized from $N_2$ under high-pressure and high-temperature conditions (130–140 GPa, >3000 K). This material crystallizes in an orthorhombic $Cmc2_1$ structure and features infinite chains with sp²-hybridized π-bonding. Theoretical calculations predict a high Tc value of 21.19 K at 113 GPa—the highest among experimentally realized non-metallic elemental materials under pressure. Remarkably, 1D-PN remains kinetically stable and energetically favorable over cg-N at ambient pressure, exhibiting an energy density as high as 8.78 kJ/g. As the first chain-like metallic nitrogen allotrope, it bridges high-energy density with electronic functionality, establishing nitrogen as a versatile system for exploring metallization and superconductivity in light elements.

**References**

1. Ji, C. et al. Ultrahigh-pressure crystallographic passage towards metallic hydrogen. *Nature* **641**, 904–909 (2025).

2. Cheng, B., Mazzola, G., Pickard, C. J. & Ceriotti, M. Evidence for supercritical behaviour of high-pressure liquid hydrogen. *Nature* **585**, 217–220 (2020).

3. Dalladay-Simpson, P., Howie, R. T. & Gregoryanz, E. Evidence for a new phase of dense hydrogen above 325 gigapascals. *Nature* **529**, 63–67 (2016).

4. Goncharov, A. F., Chuvashova, I., Ji C. & Mao, H. K. Intermolecular coupling and fluxional behavior of hydrogen in phase IV. *PNAS* **116**, 25512–25515 (2019)

5. Eremets, M. I., Hemley, R. J., Mao, H. K. & Gregoryanz, E. Semiconducting non-molecular nitrogen up to 240 GPa and its low-pressure stability. *Nature* **411**, 170–174 (2001).

6. Eremets, M. I., Gavriliuk, A. G., Trojan, I. A., Dzivenko, D. A. & Boehler, R. Single-bonded cubic form of nitrogen. *Nat. Mater.* **3**, 558–563 (2004).

7. Tomasino, D., Kim, M., Smith, J. & Yoo, C.-S. Pressure-induced symmetry-lowering transition in dense nitrogen to layered polymeric nitrogen (LP-N) with colossal Raman intensity. *Phys. Rev. Lett.* **113**, 205502 (2014).

8. Laniel, D., Geneste, G., Weck, G., Mezouar, M. & Loubeyre, P. Hexagonal layered polymeric nitrogen phase synthesized near 250 GPa. *Phys. Rev. Lett.* **122**, 066001 (2019).

9. Laniel, D. et al. High-pressure polymeric nitrogen allotrope with the black phosphorus structure. *Phys. Rev. Lett.* **124**, 216001 (2020).

10. Ji, C. *et al.* Nitrogen in black phosphorus structure. *Sci. Adv.* **6**, 920617 (2020).

11. Frost, M., Howie, R. T., Dalladay-Simpson, P., Goncharov, A. F. & Gregoryanz, E. Novel high-pressure nitrogen phase formed by compression at low temperature. *Phys. Rev. B* **93**, 024113 (2016).

12. Jiang, S. et al. Metallization and molecular dissociation of dense fluid nitrogen. *Nat. Commun.* **9**, 2624 (2018).

13. Qian, W., Mardyukov, A. & Schreiner, P. R. Preparation of a neutral nitrogen allotrope hexanitrogen $C_{2h}$-$N_6$. *Nature* **642**, 356–360 (2025).

14. Zhang, C., Sun, C., Hu, B., Yu, C. & Lu, M. Synthesis and characterization of the

pentazolate anion cyclo-$N_5^-$ in $(N_5)_6(H_3O)_3(NH_4)_4Cl$. *Science* **355**, 374–376 (2017).

15. Xu, Y. *et al.* A series of energetic metal pentazolate hydrates. *Nature* **549**, 78–81 (2017).

16. Christe, K. O. Polynitrogen chemistry enters the ring. *Science* **355**, 351–351 (2017).

17. Ashcroft, N. W. Metallic hydrogen: a high-temperature superconductor? *Phys. Rev. Lett.* **21**, 1748–1750 (1968).

18. Eremets, M. I., Drozdov, A. P., Kong, P. P. & Wang, H. Semimetallic molecular hydrogen at pressure above 350 GPa. *Nat. Phys.* **15**, 1246–1249 (2019).

19. Lipp, M. J. *et al.* Transformation of molecular nitrogen to nonmolecular phases at megabar pressures by direct laser heating. *Phys. Rev. B* **76**, 014113 (2007).

20. Liu, Y. et al. Synthesis of black phosphorus structured polymeric nitrogen. *Chin. Phys. B* **29**, 106201 (2020).

21. Zhang, J., Chen, G., Zhang, C., Xu, Y. & Wang, X. Research progress in the polymeric nitrogen with high energy density. *Chin. Phys. Lett.* **42**, 056101 (2025).

22. Mailhiot, C., Yang, L. H. & McMahan, A. K. Polymeric nitrogen. *Phys. Rev. B* **46**, 14419–14435 (1992).

23. Zahariev, F., Dudiy, S. V., Hooper, J., Zhang, F. & Woo, T. K. Systematic method to new phases of polymeric nitrogen under high pressure. *Phys. Rev. Lett.* **97**, 155503 (2006).

24. Ma, Y., Oganov, A. R., Li, Z., Xie, Y. & Kotakoski, J. Novel high pressure structures of polymeric nitrogen. *Phys. Rev. Lett.* **102**, 065501 (2009).

25. Pickard, C. J. & Needs, R. High-pressure phases of nitrogen. *Phys. Rev. Lett*. **102**,

125701(2009).

26. Wang, X. et al. Cagelike diamondoid nitrogen at high pressures. *Phys. Rev. Lett.* **109**, 175502 (2012).

27. Adeleke, A. A. et al. Single-bonded allotrope of nitrogen predicted at high pressure. *Phys. Rev. B* **96**, 224104 (2017).

28. Xinyang, L. *et al.* Prediction of $o-N_{16}$: A layered polymeric nitrogen phase. *Phys. Rev. B* **111**, 13411 (2025).

29. Wang, Y. C., Lv, J., Zhu, L. & Ma, Y. M. Crystal structure prediction via particle-swarm optimization. *Phys. Rev. B* **82**, 094116 (2010).

30. Wang, Y. C., Lv, J., Zhu, L. & Ma, Y. M. CALYPSO: a method for crystal structure prediction. *Comput. Phys. Commun.* **183**, 2063-2070 (2012).

31. Shao, X. C. et al. A symmetry-orientated divide-and-conquer method for crystal structure prediction. *J. Chem. Phys.* **156**, 014105 (2022).

32. Eremets, M. I. *High Pressure Experimental Methods* (Oxford: Oxford Univ. Press, 1996)

33. Struzhkin, V. V. et al. Superconductivity at 10–17 K in compressed sulphur. *Nature* **390**, 382–384 (1997).

34. Shimizu, K., Suhara, K., Ikumo, M., Eremets, M. I. & Amaya, K. Superconductivity in oxygen. *Nature* **393**, 767–769 (1998).

35. Fischer, N., Fischer, D., Klapötke, T. M., Piercey, D. G. & Stierstorfer, J. Pushing the limits of energetic materials—the synthesis and characterization of dihydroxylammonium 5,5'-bistetrazole-1,1'-diolate. *J. Mater. Chem.* **22**, 20418–

20422 (2012).

**Methods**

**Experimental details**

High-pressure experiments were conducted using five symmetric diamond anvil cells (DACs) equipped with 100 μm culet anvils to achieve pressures in the range of 130–140 GPa. Drilled rhenium gaskets were employed as sample chambers with hole diameters of 50–80 μm and thicknesses of 20–30 μm. A thin flake of NaCl was placed in each chamber as a thermal insulator. In experiments employing a laser-absorbing medium, a bulk absorber (Al for samples $N_2$-Al-1/2, Pb for $N_2$-Pb, and compressed TiN powder for $N_2$-TiN) occupying about one-third of the sample chamber was positioned between the NaCl layer and the sample space. High-purity nitrogen gas (99.9%) was subsequently loaded into the DACs at approximately 200 MPa using a gas loader in the lab. For comparison, one experiment (sample $N_2$-non-absorber) was performed without any absorber; in this configuration, the pressure was first raised to 140 GPa, where the compressed nitrogen itself became sufficiently absorbing to permit direct laser heating. The pressures were measured according to the edge of the diamond Raman signal. A laser beam with wavelength of 1064 nm and power of 100 W was used as the laser heating source. The laser beam was focused to a spot of 10 μm on both sides and scanned to heat the sample. Temperatures were determined from the black-body radiation of the sample. The characterizations of the samples were achieved through confocal Raman spectroscopy using the HORIBA HR EVOLUTION Raman microscope with a liquid-nitrogen-cooled CCD detector, which provide the spectral resolution of about 1 cm$^{-1}$ and the exciting laser wavelength of 532 nm. Raman spectra were measured after the samples were quenched to room temperature. The high-pressure XRD measurements of sample N$_2$-Al-1 were performed at the 16-ID-B beamline of the Advanced Photon Source (APS) using 2D mapping with a step size of 5 μm and 5 × 5 grid points. The X-ray wavelength was 0.3344 Å, and the beam size was 2 × 2 μm$^2$. Comparative analysis revealed that point 17 exhibited the most intense and numerous diffraction peaks, prompting the collection of the XRD pattern at this location.

**Computational details**

The structural predictions of solid nitrogen under high-pressure were carried out by using the CALYPSO method and first principle calculations[29-31]. Four pressures (0, 50, 100, and 200 GPa) for nitrogen under high-pressure were searched. For each pressure, fixed-cell (1, 2, 3, 4, 5, 6, and 12 formula units (f.u.) per cell) and variable-cell (1 to 12 f.u. per cell) structure searches were carried out to explore the possible phases of nitrogen under high-pressure. Based on criteria of low energy, high symmetry and structural diversity, the top 20 low-energy structures at each pressure were shortlisted from the structural predictions and subjected to high-precision reoptimization. The structural relaxations and total-energy calculations were performed using the generalized gradient approximation (GGA)[36] as implemented in the Vienna Ab initio Simulation Package (VASP)[37]. The projection-enhanced wave (PAW)[38] pseudopotentials were employed to represent the electron-ion interactions. The Perdew-Burke-Ernzerhof (PBE)[39] functionals were used to describe the exchange-correlation effects. The valence electron configuration of nitrogen atoms was set as $2s^2 2p^3$. A plane-wave cutoff energy of 900 eV was adopted. The Brillouin zone was sampled using the Monkhorst-Pack method with a dense k-point lattice of $2\pi \times 0.02$ Å$^{-1}$. The convergence criteria for energy and force were $10^{-6}$ eV/atom and $10^{-4}$ eV/Å, respectively. Raman spectra were calculated using the Quantum ESPRESSO package. Phonon dispersion curves were conducted to verify the dynamic stability using the PHONOPY code[40] based on the density functional perturbation theory (DFPT)[41]. The detonation performance and specific impulse were evaluated using the EXPLO5 (vesion6.04) and NASA CEA2 programs, respectively.

36. Perdew, J. P., Burke, K. & Ernzerhof, M. Generalized gradient approximation made simple. *Phys. Rev. Lett.* **77**, 3865–3868 (1996).

37. Vasp-Kresse,G. & Furthmüller, J. Efficient iterative schemes for ab initio total-energy calculations using a plane-wave basis set. *Phys. Rev. B* **54**, 11169–11186 (1996).

38. Paw-Blöchl, P. E. Projector augmented-wave method. *Phys. Rev. B* **50**, 17953–17979 (1994).

39. Hammer, B., Hansen, L. B. & Nørskov, J. K. Improved adsorption energetics within density-functional theory using revised Perdew-Burke-Ernzerhof functionals. *Phys. Rev. B* **59**, 7413–7421 (1999).

40. Togo, A., Chaput, L., & Tanaka I. Distributions of phonon lifetimes in Brillouin zones. *Phys. Rev. B* **91**, 094306 (2015).

41. Baroni, S., de Gironcoli, S., Dal Corso, A. & Giannozzi, P. A. Phonons and related

crystal properties from density-functional perturbation theory. *Rev. Mod. Phys.* **73**, 515–562 (2001).

**Acknowledgements**

None

**Author contributions**

K. D. conceived and directed the study. Y. L., K. D., and H. S. performed high-pressure synthesis and Raman characterization. Y. L. and A. Y. carried out X-ray diffraction experiments and structural refinement. C. L., J. M., and K. D. conducted structure searches and computational simulations. C. L., X. W., and W. Z. performed electronic structure calculations. K. D., C. L., X. W., and Z. G. jointly analyzed experimental and computational data. The manuscript was written by K. D. with contributions from X. W., C. L., A. Y., W. Z. All authors participated in scientific discussion and reviewed the manuscript.

**Competing interests**

The authors declare no competing interests.